\documentclass[iop]{emulateapj}

\slugcomment{Not to appear in Nonlearned J., 45.}

\shorttitle{Giant's Stabilizing Effect}
\shortauthors{Mu\~noz-Guti\'errez et al.}

\begin{document}


%\title{Giant Planets' Resonances and Debris Disks: 1. \\
%Global Evolution of Debris under Dwarf Planets Scattering Effects.}

\title{Giant Planets Can Act As Stabilizing Agents on Debris Disks.}


\author{M. A. Mu\~noz-Guti\'errez\altaffilmark{1}, B. Pichardo,
and A. Peimbert}
\affil{Instituto de Astronom\'ia, Universidad Nacional Aut\'onoma de 
M\'exico, Apdo. postal 70-264 Ciudad Universitaria, M\'exico}
\email{mmunoz.astro@gmail.com}


\altaffiltext{1}{Current address: UNESP - S\~{a}o Paulo State University, 
Grupo de Din\^{a}mica Orbital e Planetologia, Av. Ariberto 
Pereira da Cunha, 333, Guaratinguet\'a, SP, 12516-410, Brazil.}

%Latest update: March 13, 2017

\begin{abstract}

We have explored the evolution of a cold debris disk under the
gravitational influence of dwarf planet sized objects (DPs), both in
the presence and absence of an interior giant planet. Through detailed
long-term numerical simulations, we demonstrate that, when the giant
planet is not present, DPs can stir the eccentricities and
inclinations of disk particles, in linear proportion to the total mass
of the DPs; on the other hand, when the giant planet is included in
the simulations, the stirring is approximately proportional to the
mass squared. This creates two regimes: below a disk mass threshold
(defined by the total mass of DPs), the giant planet acts as a
stabilizing agent of the orbits of cometary nucleii, diminishing the
effect of the scatterers; above the threshold, the giant contributes
to the dispersion of the particles.

\end{abstract}

\keywords{planet-disk interactions --- protoplanetary disks --- methods: numerical}

\section{Introduction}

Apart from planets, our solar system is constituted by an enormous set
of smaller objects that go from about $2\,000$ km (dwarf planets) to
about $10^{-6}$ cm (interplanetary dust). We will refer here as a
``debris disk'', to everything within a stellar system except for the
host star and planets; we also mean for the system to be 
somewhat evolved and not forming planets anymore (i.e. not a 
prototplanetary disk); as such not much gas is expected to remain
\citep{Kral17}.
In the case of the solar system, this is mainly
the asteroid and Kuiper belts. When referring to extrasolar debris
disks, authors usually focus on the dust component (easier to
observe), instead of the larger (but less numerous) dwarf
planets. However, it has also been assumed that there should be a
spectrum of masses from dust to planetesimals that simultaneously
coexist \citep[e.g.][and references therein]
{2008ARA&A..46..339W}, as well as an effective mechanism
to continuously produce dust in these systems
\citep{1993prpl.conf.1253B,Kenyon04}.\\

The origin of debris disks is still uncertain, but it 
has been suggested 
that they might be the remnants of protoplanetary disks, 
which seem almost ubiquitous around newly formed stars
\citep{2001ApJ...553L.153H}. The masses of protoplanetary 
disks are larger
that 1M{$_\earth$} while debris disks are less massive than
1M{$_\earth$}; it is believed that the clearing processes 
of planetary disks (e.g. accretion 
to the central star and planets, photo evaporation, chaotic 
dispersion) are not totally effective, leaving residual solid
material behind that may be the origin of debris disks 
\citep{2008ARA&A..46..339W}.\\

The dust in
debris disks could be either generated as the part of the original
dust component of the protoplanetary disk that was not cleared up
(a.k.a. primordial dust), or,
assuming the clearing processes 
also left behind large planetesimals
($\ga$ 1 km), produced by planetesimal collisions; the time scales
observed for the older debris disks supports this last scenario
\citep{2002MNRAS.334..589W}, although for the younger ones, it might
be directly the dust left behind by protoplanetary disks.\\

Debris disks also seem to be common around young objects and 
their presence sometimes lasts for as long
as the planetary system has existed. The evolution of debris disks may
be influenced by planet formation processes, which 
last for about 1 Gyr,
the time required by the system to settle into a stable
configuration \citep{2008ARA&A..46..339W}.\\

The first extrasolar debris disk was discovered from the thermal
emission by dust observed with the IRAS satellite in the star Vega
\citep{1984ApJ...278L..23A}. Since that time, about $1\,750$ debris
disks have been found by their dust emission 
\citep[e.g.][]{1992A&AS...96..625O,1998ApJ...497..330M,
2016ApJS..225...15C}, we also expect a large quantity of dust-poor
debris disk systems, similar to the Kuiper belt
\citep{2008ARA&A..46..339W,Moro08,Nesvorny10,Vitense12}.\\

The morphological and dynamical characteristics of debris disks are
influenced by the configuration of the hosting planetary system
\citep[e.g.][]{1997MNRAS.292..896M,1999ApJ...527..918W,2014prpl.conf..521M,2015ApJ...798...83N,2016ApJ...827..125L,2016ApJ...826...19N,2017arXiv170206578N};
in particular, the formation of giant planets has an important
dynamical effect on protoplanetary disks. Once one or more 
planets has formed and the gas has dissipated, the planets will 
continue to gravitationally influence the debris disk. There are important
effects known by the presence of a planet close to a debris disk, such
as: stirring through secular and resonant perturbations, resonant gaps
\citep{2008ARA&A..46..339W,Mustill09}, among others. A nearby example is the
case of the Kuiper belt that shows evidence of sculpting due to the
migration of Neptune \citep{2007prpl.conf..895C,2008ssbn.book..275M,
Nesvorny15};
all these works consider the giant planet as an orbital ``disorderer''
agent.  
In this work we are quantifying an effect presented in
\cite{Munoz15}; specifically we demonstrate that giant planets, 
depending on the
total mass of a given debris disk, can also act as a stabilizing
mechanism on the cometary nucleii, and it can quench the dynamical
evaporation induced by the presence of DPs on debris disks 
.\\

This paper is organized as follows: in Section \ref{sims} we provide
the details of the numerical simulations we performed; we also 
describe 
the distributions of DPs, the debris disk and the giant planet used in
this work. In Section \ref{res} we present and discuss the results of
our simulations, while in Section \ref{conc} we discuss and enumerate
the main conclusions of this work.\\

\section{Simulations}
\label{sims}

We perform a large set of simulations of debris disks composed of 50
DPs and $5\,000$ massless test particles which represent cometary
nucleii. All of our simulations were performed using the mixed
symplectic integrator of the MERCURY package \citep{Chambers99} with
an initial time-step of 180 days and an accuracy parameter for the
Bulirsh-St\"oer integrator of $10^{-10}$, which solves the evolution
of test particles when their distance to any major body is smaller
than 3 Hill radius ($R_{\rm H} = a_p(M_p / 3 M_{\odot})^{1/3}$; where
$a_p$ and $M_p$ are the semimajor axis of the orbit and the mass of
the planet, respectively).  Most of our simulations are 1 Gyr long but
for some cases we extended the simulations to 2 Gyr.

\subsection{Test particles's initial conditions: Random Cold 
Debris Disk}\label{tpd}

Taking as a reference our own solar system (or more precisely, the
Kuiper belt), we generate a random belt of debris distributed between
38 and 50 AU in semimajor axis, $a$, which consists of $5\,000$ test
particles. The direction of the angular momentum is determined as a
2-dimensional Gaussian distribution centered in the Z-axis with a
width of $1.2^\circ$ in each, X and Y axes \citep[see the description
  in][]{Munoz15}; this translates to an average inclination of
$\mu_i=1.47^\circ$, an inclination dispersion of
$\sigma_i=0.75^\circ$, and at the same time defines a homogeneous
random distribution of the longitude of the ascending node, $\Omega$,
between 0 and 360$^\circ$.  The particle eccentricity, $e$, and the
argument of pericenter, $\omega$, are defined similarly: first
randomly assigning the locus of the second focus of each elliptical
orbit (the first being the Sun) with a Gaussian probability
distribution in the XY plane, with a center coinciding with the Sun
and a width of $0.03$ along each axis; this translates to an average
eccentricity, $\mu_e=0.037$, an eccentricity dispersion,
$\sigma_e=0.02$, and at the same time defines a homogeneous random
distribution of $\omega$ between 0 and 360$^\circ$. Finally, the mean
anomaly, $M$, is randomly generated between 0$^\circ$ and
360$^\circ$.\\

This disk, with the same random seeds, is used in all our 
simulations as the test-particle/cometary-nucleii initial 
conditions.\\

\subsection{Dwarf Planet Distributions}
\label{sec:dpdis}

Our simulations require a suitable number of scatterers embedded in
our cold belt in order to be both a) perturbative enough in the
presence of a giant planet and b) computationally fast enough to 
run the required number of simulations in a feasible amount of 
time. From the experiments performed in
\citet{Munoz15} we consider that 50 scatterers are adequate to fulfill
both requirements. The 50 DPs we use represent the 50 largest DPs in
such debris disk. In the rest of this work we use different
distributions of 50 DPs each, which differ from each other by the
maximum inclination of the set, the total mass of the DPs, and the
index of their differential mass distribution, $\alpha$.  Below we
describe the different DP distributions employed.\\

First, we used a differential mass distribution (dMD) derived from the 
differential size distribution (dSD). A dSD is given by
\begin{equation}
\frac{dN}{dD}\propto D^{-\alpha'},
\end{equation}
where $N$ is the number of objects larger than $D$.
It has been shown that $\alpha'$ values range from 3.5, for belts in 
collisional equilibrium
\citep{Dohnanyi69},
to 4.0 or 4.5, for the largest objects in the 
Kuiper Belt \citep{Fraser08a,Fraser08b}. It is
straightforward to obtain a dMD considering that for a constant
density $D\propto M^{1/3}$;
thus a dMD will be given by
\begin{equation}\label{dMD}
\frac{dN}{dM}\propto M^{-\alpha},
\end{equation}
where $N(M)$ is the number of objects more massive than $M$ and where
$\alpha=(\alpha'+1)/3-1$ takes values from $\sim$1.8, for disks in
collisional equilibrium, to $\sim$2 or $\sim$2.2, for the largest
Kuiper belt objects.\\

In order to compute a random mass distribution of DPs that follows
equation \ref{dMD}, for each $\alpha$ we assigned the mass of the n-th
planet as:
\begin{equation}\label{massdist}
m_n=\left(\frac{K_{\alpha}}{(\alpha-1)n'}\right)^{1/(1-\alpha)}.
\end{equation} 
Here $n'$ is a random real number in the interval $[0.5,50]$, 
while $K_{\alpha}$ 
is a constant related with the largest mass allowed
in the distribution of DPs. We set the maximum value of $n'$ to 
50 to represent the 50 heaviest planets of our sample; in a more 
realistic simulation we would expect to have 50 more objects with 
$50<n'<100$ (and even a greater number of smaller objects), 
however these objects 
are expected to have a much smaller contribution to the dynamics 
of the disk; we set the minimum value of $n'$ at 0.5 to avoid the 
possibility of a single very large planet to completely dominate 
the dynamics of the disk; $K_{\alpha}$ is set so the total mass of
the 50 DPs equals $(\sqrt{5})^q\times M_{\it CKB}$, where 
$M_{\it CKB}$ is the 
estimated mass of the classical Kuiper belt, set as 
$M_{\it CKB}=0.01$M$_\oplus$
\citep{Bernstein04,Fraser14}, while $q$ is an integer in the interval 
$[0,4]$; this covers
the cases where the addition of the masses of the 50 DPs equal 
0.01, 0.0223, 0.05, 0.1118, and 0.25 M${_\oplus}$; 
therefore, $K_{\alpha}$ varies for each index $\alpha$ 
considered and for the total mass of DPs.  
For example, for an index
$\alpha=2$ and a total mass of 0.01 M${_\oplus}$, $K_{\alpha}$ will be
of the order of the mass of Pluto, while the heaviest possible planet
in this distribution would have twice Pluto's mass. It should be 
noted that in our simulations, the total mass of
the debris disk corresponds to the sum of the masses of the 50 DPs.\\

\begin{figure}
\plotone{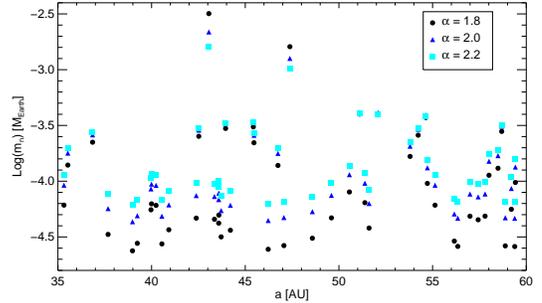}
\caption{Distribution of DPs for different dMD indexes. The mass for
  each DP in the distribution is plotted against its semimajor axis,
  for three different distributions of DPs with total mass
  0.01M${_\oplus}$ but with different $\alpha$ indexes, as indicated
  in the figure.}
\label{avsm}
\end{figure}

As an example, we show in Fig. \ref{avsm} the masses and positions, in
semimajor axis, of three distributions of DPs; they differ in the
index of their dMD but their total mass equals 0.01M${_\oplus}$ in all
cases. This distribution is general throughout this work in the sense
that the position in $a$ will be the same, this since the random seed
we use to define the DPs is the same for all simulations.  Note,
however, that the individual mass of each DP will be rescaled;
although the sets of $m_n$ will not be identical, the sets of $n'$s
will be identical, which will, using equation \ref{massdist}, produce
the sets of $m_n$ (for the $\alpha$ and $K_\alpha$ that correspond to
each simulation).  The particular $a$ distribution generated in this
work has the advantage that the two heaviest DPs lie almost in the
middle of the debris disk, thus their effect is highlighted, while the
smallest DPs are randomly exerting their influence along the disk and
beyond its limits.\\
 
Regarding the maximum possible inclination of DPs, $i_{max}$, we
explore 3 different cases: cold distributions have $i_{max}=5^\circ$,
warm distributions have $i_{max}=15^\circ$, and hot distributions have
$i_{max}=30^\circ$. For each case inclinations of DPs are assigned
randomly between 0 and $i_{max}$.\\

The remaining orbital parameters are the same for all distributions as
follows: $a$ is randomly determined from a uniform distribution
between 35 and 60 AU, $e$ is randomly distributed between 0 and 0.1,
and finally $\omega$, $\Omega$, and $M$ are set randomly between
0$^\circ$ and 360$^\circ$.\\

\subsection{Giant Neptune-like Planet}

\begin{table}
\centering
\caption{Giant planet's physical and initial orbital parameters.}
\label{tab:tabNep}
\begin{tabular}{@{}lc}
\hline
\hline
Constant & Value \\
\hline
GM (km$^3$ s$^{-2}$) & 6.836534062$\times10^6$ \\
Radius (km) & $24\,043$ \\
$a$ (AU) & 30.09598340353801 \\
$e$ & 0.01078167440459 \\
$i$ ($^\circ$) & 0.0 \\
$\omega$ ($^\circ$) & 272.2034594991346 \\
$\Omega$ ($^\circ$) & 131.7818550161205 \\
$M$ ($^\circ$) & 288.9806490025501 \\
\hline
\end{tabular}

\medskip
\end{table}

The influence of DP distributions over the debris disk are tested with
and without the presence of an interior giant planet; such giant
shares all the orbital and physical characteristics of Neptune, except
for its nominal inclination.  The kinematics of the outer solar system
are dominated by Neptune, in particular any debris disk should align
itself with Neptune's inclination.  The same will happen with any
other debris disk with an interior giant planet. To avoid an initial
rearrangement of test particles, we set the giant inclination to
0$^\circ$, instead of 1.8$^\circ$ as defined in the solar system.\\

Table \ref{tab:tabNep} shows the physical parameters of the
Neptune-like planet used in this work, as well as its initial orbital
osculating parameters for the simulations. A central star of 1
M$_\odot$ is present in all simulations.\\

Table \ref{tabsim} shows a summary of all parameters we can vary in
each simulation. There are 90 different scenarios resulting from
combinations of these parameters, but we have not run all possible
cases. In what follows, when needed, we will refer a particular
simulation with the generic name $|GP_{(Y/N)}/\alpha\, x/ i_{max}\, y/
M_{DP}\, z|$, where $GP_{(Y/N)}$ refers to the inclusion or not of the
giant planet in the simulation, and $(x,y,z)$ are the possible values
of each varied parameter. For example: a simulation including the
giant planet, with a distribution of DPs with a dMD index $\alpha=2$,
a maximum inclination $i_{max}=5^\circ$, and a total mass of 0.05
M$_{\oplus}$, will be labeled $|GP_{Y}/\alpha\, 2/ i_{max}\, 5/
M_{DP}\, 5|$.\\

\begin{table}
\caption{Summary of varied parameters in different simulations.}
\label{tabsim}
\begin{tabular}{@{}lccccc}
\hline
Parameter & Value & & & \\
\hline
GP & Yes & No & & & \\
$\alpha$ & 1.8 & 2.0 & 2.2 & & \\
$i_{max}$ & $5^{\rm{o}}$ & $15^{\rm{o}}$ & $30^{\rm{o}}$ & & \\
$M_{DPs}$ & 1.0 & 2.23 & 5.0 & 11.18 & 25.0 \\
\hline
\end{tabular}

\medskip
Here $M_{DPs}$ is given in units of masses of the classical Kuiper
belt, $M_{\it CKB}=0.01$M$_\oplus$.
\end{table}

\section{Results}
\label{res}

Dwarf planet sized objects have been commonly underestimated when
considering the dynamical evolution of debris disks. Those massive
bodies can be considered as the far end of the size
distribution in protoplanetary and debris disks. In this work we show
the importance of dwarf-planet sized objects in the secular evolution
of a cold debris disk; we do this as a function of the total mass in
DPs and other critical orbital characteristics.\\

To analyze the changes in the global properties of our debris disk, we
have measured the evolution of the average orbital elements,
$\left<e\right>$ and $\left<i\right>$, as well as their associated
standard deviations, $\sigma_e$ and $\sigma_i$, for the whole
population of the disk. We have not shown the evolution of
$\left<a\right>$ because, in general, this parameter does not evolve
significantly, thus, it can not provide us with valuable information.
Of course, some particles will experience a significant 
change in semimajor axis due to perturbations by the DPs, but in
average, the position of the disk remains unchanged. \\

On the other hand, both $\left<e\right>$ and $\left<i\right>$, are
related to the dynamical heating and the thickness of the disk; this
because of the maximal excursions of particles above the reference
plane, related to $\left<i\right>$, and in the radial direction,
related to $\left<e\right>$.\\

\subsection{Effect of Heavy End of the Mass Distributions}

As we mentioned earlier in Sec. \ref{sec:dpdis}, we have tested three 
possibilities for the 
dMD index of DPs, where the lower
limit, $\alpha=1.8$, relates to the heavy end of the size 
distribution of a debris disk in collisional equilibrium, while the
upper limit, $\alpha=2.2$, relates to the sizes of the largest objects
in the Kuiper belt, which constitutes our reference system. \\

\begin{figure*}
\plotone{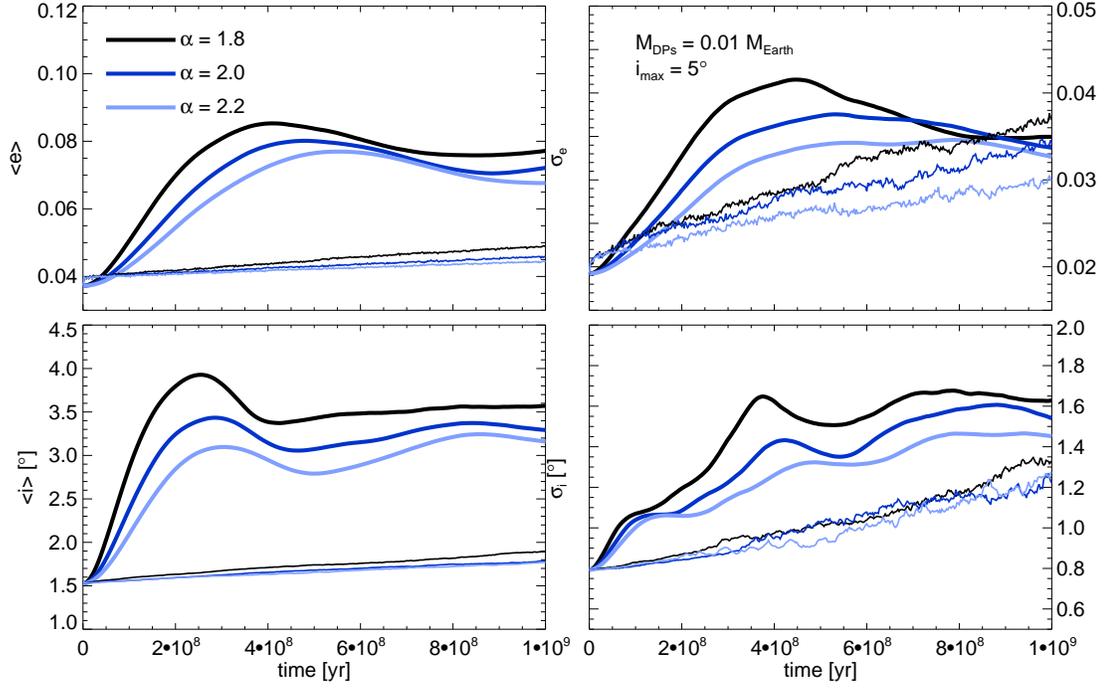}
\caption{Evolution of $\left<e\right>$, $\sigma_e$, $\left<i\right>$, 
and $\sigma_i$ of the particles of debris disks with 
$M_{DPs}=0.01$M$_\oplus$ and 
$i_{max}=5^\circ$. Black lines represent disks with $\alpha=1.8$, dark 
blue lines disks with $\alpha=2.0$, and light blue lines disks with
$\alpha=2.2$; thick lines are for simulations without the giant planet,
while thin lines are simulations that included the Neptune-like
giant.}
\label{ASAlph}
\end{figure*}

Figure \ref{ASAlph} shows the evolution of $\left<e\right>$ (left
upper panel), $\left<i\right>$ (left bottom panel), $\sigma_e$ (right
upper panel), and $\sigma_i$ (right bottom panel) for 6 simulations
that share two parameters: $i_{max}=5^{\circ}$ (a cold population of
DPs), and their total mass, $M_{DPs}=0.01$M$_\oplus$.  Three
simulations included the giant planet (shown in thin lines), while the
other three did not (shown in thick lines); we used three different
indexes $\alpha$ in each case: black lines are simulations in which DP
distribution follows a dMD with $\alpha=1.8$, blue lines for
$\alpha=2.0$, and light blue lines for $\alpha=2.2$. \\

Evident from Fig. \ref{ASAlph} is the weak dependence of the global
evolution of the disk on the parameter $\alpha$ used to generate the
distribution of DPs. Nonetheless, it is worth to note that for
simulations without the giant planet, the $\alpha=1.8$ distribution
(thick black lines) leads to a higher global stirring of the orbits
than the $\alpha=2.2$ case (thick light blue lines), while the
$\alpha=2.0$ (thick blue lines) lies in the middle. Now, if we recall
our $m_n$ {\it vs} $a$ distribution (Fig. \ref{avsm}), we note that
the two more massive DPs for a fixed total mass belong to the
$\alpha=1.8$ distribution. This fact alone tells us that just the two
more massive bodies can dominate the heating of the debris disk,
despite the fact that most of the less massive bodies of the same
$\alpha=1.8$ distribution are less massive than the corresponding
objects in the $\alpha=2.2$ distribution (i.e. the presence 
of larger-mass perturbers excites larger eccentricities and 
inclinations in the disk, regardless of the presence of lower-mass 
perturbers.).\\

Another general feature from Fig. \ref{ASAlph}, for the simulations
which do not include the giant planet, is the initial bump evident in
both $\left<e\right>$ and $\left<i\right>$. This bump results from the
initial shock given by DPs to the disk, before the particles reach a
relaxed state with the DPs generated potential. After such relaxation
the population continues its $\left<e\right>$ and $\left<i\right>$
evolution in a slower and steady pace. This is a typical behavior for
the evolution of the whole disk in all our simulations.\\

It is worth to note that, without the giant planet, both the maximum
values of $\left<e\right>$ and $\left<i\right>$, reached at the top of
the initial bump, remain below the maximum $e$ and $i$ of the DP
distributions (0.1 and 5$^\circ$, respectively). From this conclude
that the lowest mass explored in this work, approximately the mass of
the classical Kuiper belt, is not enough to excite the average
particle eccentricity/inclination up to the excitation level of the
DPs, not at the bump nor at any other point of the 1 Gyr simulation
(we estimate they would require 10 Gyr or more to reach this
level of excitation).\\
 
An important result steaming from Fig. \ref{ASAlph} is the damping
effect that a Neptune-like giant planet produces on the stirring of
the inclinations and eccentricities of the disk particles, which
otherwise would be increased by the perturbations of DPs
alone. Indeed, the values reached by $\left<e\right>$ and
$\left<i\right>$ at the end of the simulation, owing just to the DPs'
influence are $\left<e\right>\sim0.07$ and $\left<i\right>\sim3.3^{\rm
o}$, while, when the giant planet is included, the maximum values
reached are $\left<e\right>\sim0.045$ and $\left<i\right>\sim1.8^{\rm
o}$, for the 3 different dMD indexes.  This effect was previously
pointed out by \citet{Munoz15}, noting that an interior giant planet
of small eccentricity, seems to stabilize the orbits of debris disk
particles, preventing an important stirring produced by DPs. Later on,
we will discuss the origin of this important phenomenon. However, from
this first experiment we can see the weak dependence of such effect on
the dMD of DPs: since for the three indexes the damping is of the same
order and the relative behavior, between different distributions of
DPs, is maintained in both $\left<e\right>$ and $\left<i\right>$.\\

\subsection{Effect of Progressively Hotter DPs Populations}

\begin{figure*}
\plotone{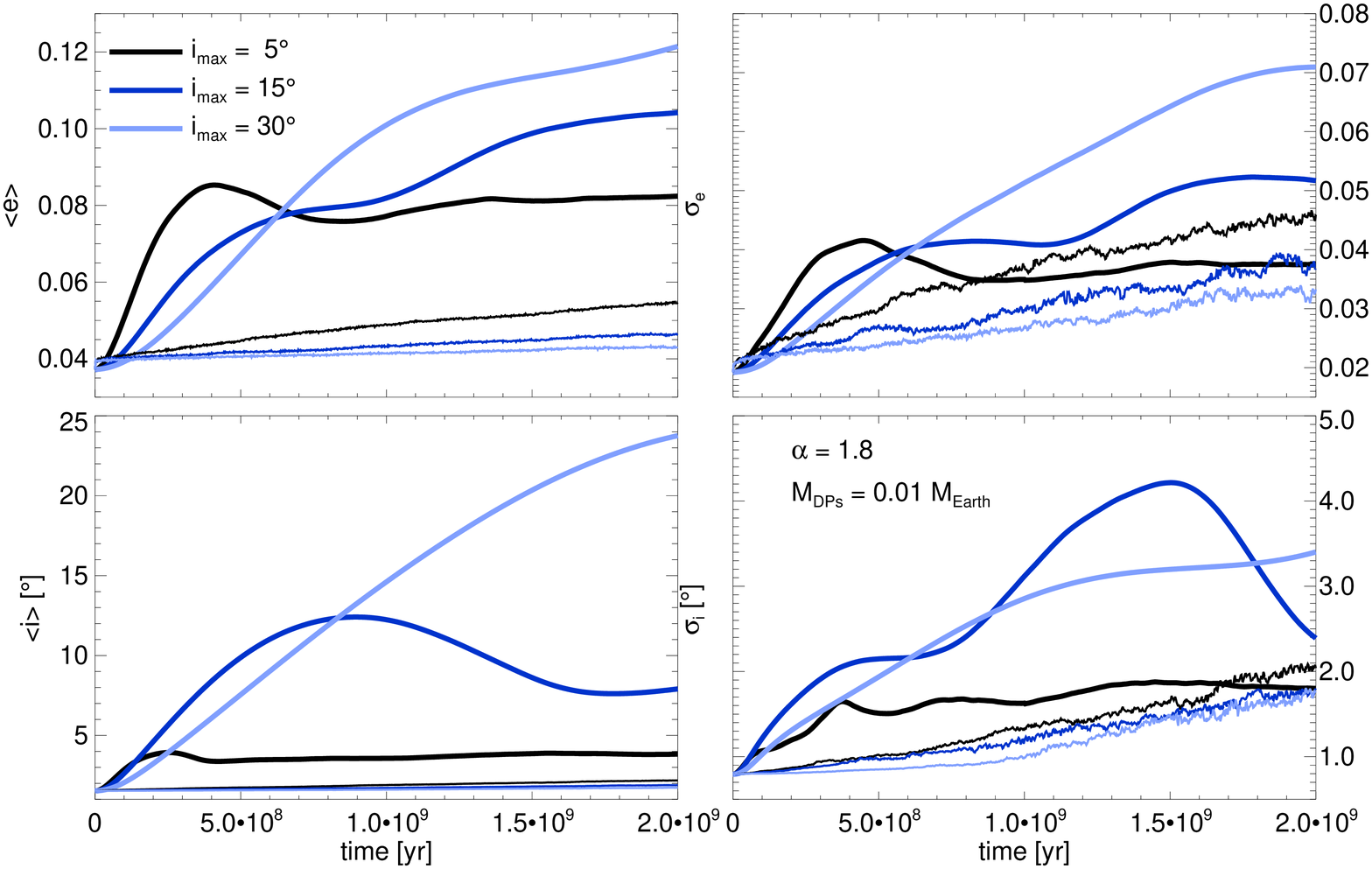}
\caption{Evolution of $\left<e\right>$, $\sigma_e$, $\left<i\right>$,
and $\sigma_i$ of the particles of debris disks with
$M_{DPs}=0.01$M$_\oplus$ and $\alpha=1.8$. Black lines represent
cold disks with $i_{max}=5^\circ$, dark blue lines warm disks with
$i_{max}=15^\circ$, and light blue lines hot disks with
$i_{max}=30^\circ$; same as in previous figure, thick and thin lines
are for simulations without and with the giant planet,
respectively.}
\label{ASInc}
\end{figure*}

Fig. \ref{ASInc} shows the evolution of the global disk quantities
$\left<e\right>$ and $\left<i\right>$, as well as their standard
deviations $\sigma_e$ and $\sigma_i$; this time the results stand for
simulations that include DP populations of different vertical
excitations. We label the populations as a function of their maximum
initial inclination, we label cold distributions (in black lines) when
$i_{max}=5^\circ$, warm distributions (blue lines) for
$i_{max}=15^\circ$, and hot distributions (light blue lines) when
$i_{max}=30^\circ$. Again we differentiate the cases including the
giant Neptune-like planet (in thin lines) from those without it (in
thick lines).  The total mass of all DP distributions in this case is
$0.01$M$_\oplus$ and $\alpha=1.8$ in all simulations shown in this
section.\\

We extend this simulations to 2 Gyr (which is further than the 1 Gyr
used in the previous section) to better see the evolution of the
initial bump produced by the DPs and the subsequent relaxation and
steady evolution of the debris disk particles. This
relaxation is evident in the thick black curves of Fig. \ref{ASInc};
one can imagine that the thick dark blue curves are near relaxation,
while for the other 4 simulations it is evident that much more time
is needed to reach relaxation.\\

In \citet{Munoz15}, we had speculated about the maximum average
inclination reached by the disk particles in our simulations as a
function of the maximum initial inclination of the DP
distributions. In Fig. \ref{ASInc} we see that this could be indeed
the case when the disk evolves without the presence of the giant
planet (thick lines), i.e., larger initial DP inclinations imply a
greater vertical heating of the disk. Nevertheless, some subtleties
should be noted: first, the maximum $\left<i\right>$ reached at the
top of the initial bumps in all cases, is always below the $i_{max}$
of the corresponding distribution; even if the top of the bump
is not reached in 2 Gyr for the hot distribution of DPs, its
curvature already shows the maximum will likely be below 30$^\circ$
(maximum of the distribution). On the other hand, the warm and hot
distributions can increase the maximum $\left<e\right>$ of the test
particles beyond the maximum $e$ of the DPs (set in all cases to 0.1),
this is not the case for the cold distribution where $\left<e\right>$
settles near 0.08.\\

Second, the slopes of black, blue, and light blue thick curves in the
left and right panels of Fig. \ref{ASInc} are progressively less
steep, which means that, as the DP inclinations become larger, their
effect will be slower but, in the long run, it will be greater. This
is evident in $\left<i\right>$ and $\sigma_i$ (Fig. \ref{ASInc},
bottom panels), but also clear in $\left<e\right>$ and $\sigma_e$
(Fig. \ref{ASInc}, upper panels). With time, in distributions without
the giant planet, the scattering produced by hotter DP distributions
on a cold disk will be larger, but the timescale required for this
will also be larger. The latter behavior is to be expected as the
relative velocities of individual interactions between highly inclined
DPs and particles in a cold disk will be greater, and consequently the
perturbation experienced by the cometary nucleii will be quicker and
therefore smaller (or equivalently, when the DP distribution
is hotter, the DPs will spend less time near the disk, which means
there will be less close encounters when they are hotter). As a
confirmation of this point we note that the previously mentioned
initial bump produced by DPs and the subsequent relaxation of the
particles is clearly visible in the evolution of $\left<i\right>$
(left bottom panel of Fig. \ref{ASInc}) for the $|GP_{N}/\alpha\, 1.8/
i_{max}\, 5/ M_{DP}\, 1|$ and $|GP_{N}/\alpha\, 1.8/ i_{max}\, 15/
M_{DP}\, 1|$ simulations, but not for the $|GP_{N}/\alpha\, 1.8/
i_{max}\, 30/ M_{DP}\, 1|$, which even after 2 Gyr has not reached the
relaxation expected after the initial bump.\\

The inclusion of the giant planet modifies significantly the
behavior. First, as already pointed out, the giant damps the
scattering produced by DPs (regardless of their maximum inclinations)
and second, the stirring produced by DPs is weaker for larger
inclinations, contrary to the case when only DPs are present. Note the
inverse relation between the thin and thick lines, at the end of the
simulations in all panels of Fig. \ref{ASInc}. This means that hotter
distributions of scatterers will be less effective in heating a cold
disk if a stabilizing interior giant planet is present.\\

From Fig. \ref{ASInc} it can be seen that the black curves are
initially steeper than the blue ones, but at a certain time they
saturate and stop growing, earlier than less steep curves; in
the case of $\left<i\right>$, the expected peak for each
distribution is directly related to the $i_{max}$, therefore,
eventually the light blue line should surpass the black one;
however, this would take over 100 Gyr, much longer than the expected
lifetime of the system. In other words, in the presence of the giant
planet the heating produced by hotter DP distributions would
required a huge amount of time. For these simulations, it can be
said that the effect of the giant only delays the stirring produced by
DPs, but for the timescales involved one can safely argue that the
giant planet is effectively a stabilizing agent of the debris disk.\\

\subsection{Effect of Increasing the Total Mass of DPs}

\begin{figure*}
\plotone{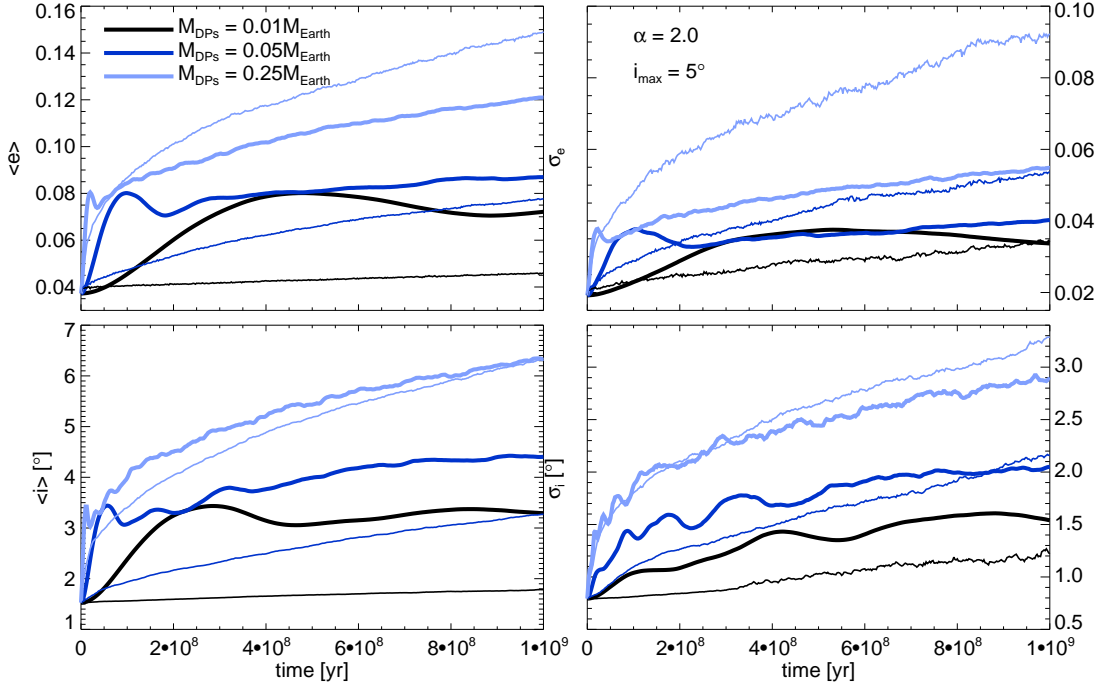}
\caption{Evolution of $\left<e\right>$, $\sigma_e$, $\left<i\right>$,
and $\sigma_i$ of the particles of cold debris disks with
$i_{max}=5^\circ$ and $\alpha=2.0$. Black lines represent disks with
$M_{DPs}=0.01$M$_\oplus$, dark blue lines disks with
$M_{DPs}=0.05$M$_\oplus$, and light blue lines disks with
$M_{DPs}=0.25$M$_\oplus$; thick lines for simulations without giant;
thin lines for simulations including the giant.}
\label{ASMass}
\end{figure*}

In the simulations shown in Figure \ref{ASMass} we varied the total
mass of DP distributions; we present three cases: $0.01$M$_\oplus$
(black), $0.05$M$_\oplus$ (blue), and $0.25$M$_\oplus$ (light
blue). In Table \ref{tabsim} these masses correspond to 1, 5, and 25
$M_{\it CKB}$.  As before, simulations that include the giant planet
are shown in thin lines while simulations without the giant are shown
in thick lines. The constant parameters in these simulations are the
maximum inclination of DP distributions, set to $i_{max}=5^{\rm o}$
and the dMD index, set to $\alpha=2$.\\

Without the giant planet (thick lines), we observe that, as the DPs
become more massive, their effect on increasing $e$ and $i$ becomes
more rapid (without increasing the initial bump); 
raising $\left<e\right>$ and
$\left<i\right>$ to values of $\sim0.08$ and $\sim3.5^{\rm o}$
respectively, at the top of the initial bump, regardless of the
total mass of the scatterers. This implies that the initial
rearrangement of the particles will respond to the initial DP's $e$,
and $i$ distributions and not to their mass, but this rearrangement
occurs sooner for more massive DPs. After the initial bump, later
steady evolution is smooth, both in $\left<e\right>$ and
$\left<i\right>$. As one would expect, the more massive the DPs the
more scatter they will produce, over a set amount of time, on the
disk.\\

Additionally, simulations $|GP_{N}/\alpha\, 2/ i_{max}\, 5/ M_{DP}\,
1|$ and $|GP_{N}/\alpha\, 2/ i_{max}\, 5/ M_{DP}\, 5|$ seem to reach a
saturation limit, i.e., a maximum $\left<e\right>$ and
$\left<i\right>$ after which the DPs can no longer increase the
average values of the population; on the contrary, simulation
$|GP_{N}/\alpha\, 2/ i_{max}\, 5/ M_{DP}\, 25|$ shows a monotonically
growing $\left<e\right>$ and $\left<i\right>$ evolution, even after
$\left<i\right>$ reaches the $i_{max}\sim5^\circ$ limit that we have
previously suggested as the saturation for a distribution of cold DPs.
This means that massive enough DPs can stir particle inclinations at
least above their own maximum limit.\\

Now lets consider the simulations including the giant planet.  For the
0.01M$_\oplus$ DP distribution the result from including the giant
planet is similar to what is seen in the black lines of
Fig. \ref{ASInc}, i.e., the heating suppression due to the giant. But,
for the most massive DP distribution (0.25M$_\oplus$), this behavior
reverses, and now, at the end of the simulation, the scattering
produced by DPs plus the giant is larger than the one produced by DPs
alone, as one would expect from adding mass to the scattering
population. This behavior is evident in light blue curves for
$\left<e\right>$, $\sigma_e$, and $\sigma_i$ of Fig. \ref{ASMass},
where the thin lines rapidly surpass the thick ones after just
$\sim$50 Myr, for the cases of $\left<e\right>$ and $\sigma_e$, and
after $\sim$300 Myr, for $\sigma_i$. Although slower, this also seems
to occur for $\left<i\right>$, near the 1 Gyr limit of our
simulations. \\

This result shows that a stabilizing 
mass-related threshold exists; this threshold probably 
depends on the mass of the
giant planet (the larger the mass of 
the giant, the larger the mass of the disk it can stabilize). \\

For a fixed mass of
the giant planet, the threshold depends on the total mass of the
disk; below this threshold
a massive interior planet is able to slow down the heating 
of cometary nucleii due to
the action of massive dwarf-sized scatterers, but above such threshold, 
the giant contributes to the dispersion of the orbits. From 
Fig. \ref{ASMass} we estimate this threshold to be between 
0.05 and 0.25 M$_\oplus$
for the scatterers considering a Neptune-mass giant.\\

This implies that, in the presence of a giant planet, very massive 
disks will quickly evaporate, but along with the cometary nucleii, 
many DPs will also disperse until the mass of the disk reaches 
the threshold value when the evolution will slow down. We 
indeed see an increase on the loss of cometary nucleii in the more
massive simulations. Besides, when the giant planet is included, at least one 
DP is lost in every simulation, and even two in the simulation 
$|GP_{Y}/\alpha\, 1.8/ i_{max}\, 15/ M_{DP}\, 1|$, when their semimajor
axes grow larger than 100 AU.\\ 

\subsection{A Mass Threshold for the Stabilizing Effect of the Giant}

The mass threshold we aim to determine
depends on the mass of the
scatterers (DPs), the mass of the giant planet 
as well as its distance to the disk. In the present work
we fixed the mass of the giant planet and its distance to
the disk (Neptune-like).  
Additionally, the giant's eccentricity and inclination could have and
important effect, nonetheless
we left the exploration of
other parameters for a later work. In this paper we focus on the 
existence of two regimes, the first one, when the giant planet
acts as a stabilizing agent of disk particles, and the second, when the 
giant also contributes to
the dispersion of particles in the disk, once its ordering influence 
over cometary nucleii is surpassed by DP perturbations.\\

\begin{figure}
%\epsscale{.80}
\plotone{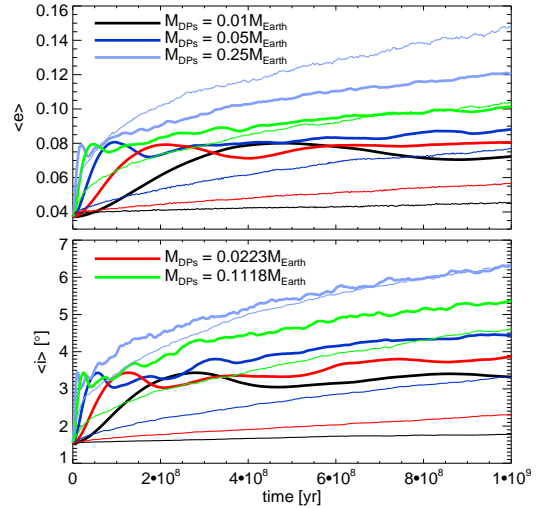}
\caption{Evolution of $\left<e\right>$ and $\left<i\right>$ of the
particles of cold debris disks with $i_{max}=5^\circ$ and
$\alpha=2.0$. As in previous figures, black lines represent disks
with $M_{DPs}=0.01$M$_\oplus$, dark blue lines disks with
$M_{DPs}=0.05$M$_\oplus$, and light blue lines disks with
$M_{DPs}=0.25$M$_\oplus$; two new cases for disks with
$M_{DPs}=0.0223$M$_\oplus$ and $M_{DPs}=0.1118$M$_\oplus$ are shown
in red and green lines, respectively. Again thick lines are for
simulations without the giant planet; thin lines for simulations
including the giant.}
\label{massth}
\end{figure}

The two regimes we find can be guessed already from the results shown
in Fig. \ref{ASMass}, mainly those on the top-left panel, where the
evolution of $\left<e\right>$ is plotted for 3 different cases which
differ by the total mass of the DP distributions. In such figure there
is a noticeable change in the behavior of $\left<e\right>$ evolution;
when the total mass is 0.05M$_\oplus$, the thin dark blue curve
remains below the thick dark blue curve, this means that the effect of
the giant still results in the stabilization of cometary nucleii,
while when the mass of DPs is 0.25M$_\oplus$, the thin light blue
curve rapidly surpasses the thick light blue one, i.e. in such case
the giant contributes to the heating of the disk. Therefore, the
change of regime occurs between the 0.05 and 0.25M$_\oplus$. To
account for the gaps in the intermediate masses and better cover the
full mass interval (0.01 to 0.25M$_\oplus$), we added the cases for
total masses of 0.0223 and 0.1118M$_\oplus$. For this purpose we ran 1
Gyr integrations of $2\,000$ test particles in the debris disk. We
used only cold distributions of DPs for the cases with and without the
Neptune-like giant (Table \ref{tabsim} lists the total DP masses
covered in this experiment).\\

Fig. \ref{massth} shows the evolution of $\left<e\right>$ and
$\left<i\right>$ for the 5 DP distributions, with and without the
giant planet. We plotted again as black, dark blue, and light blue
curves, the evolution of $\left<e\right>$ and $\left<i\right>$ for the
0.01, 0.05, and 0.25M$_\oplus$ cases, respectively (those shown in
Fig. \ref{ASMass}), but now including only $2\,000$ cometary nucleii
in the disk. Reducing the number of test particles from $5\,000$ to
$2\,000$ has not modified significantly the statistical evolution of
the disk as a whole, as can be seen by comparing left panels of
Fig. \ref{ASMass}, with the black and blue curves, both thick and
thin, of Fig. \ref{massth}. The newly cases (0.0223 and
0.1118M$_\oplus$) are plotted in different color to highlight the
completeness they provided in covering the full mass interval.  In red
and green, we plotted the evolution of $\left<e\right>$ and
$\left<i\right>$ for the 0.0223 and 0.1118M$_\oplus$ cases,
respectively.\\

The anticipated behavior of both $\left<e\right>$ 
and $\left<i\right>$ is observed for the cases plotted in red and green. 
In red, thick
and thin lines (with and without the giant) remain between the black and 
dark blue curves, while the green lines remain between the dark blue
and light blue lines.\\

\begin{figure}
%\epsscale{.80}
\plotone{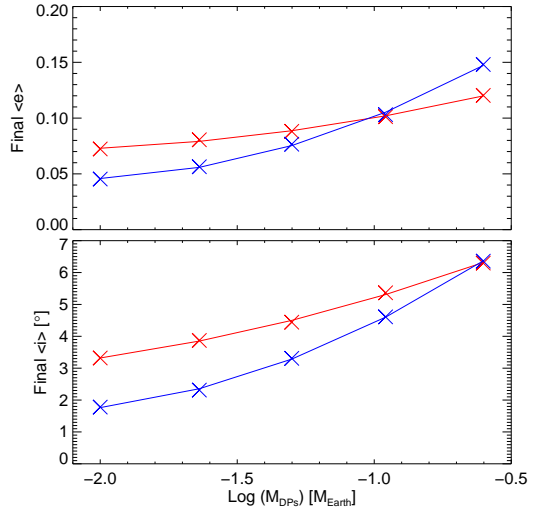}
\caption{Final values of $\left<e\right>$ and $\left<i\right>$ 
for the different masses of DP distributions used. Red crosses
stands for simulations without the giant planet, while blue crosses
are for simulations that included the Neptune-like giant. Second order
fittings are used to find the mass where the giant changes of
regime, from stabilizing to disperse particles in the disk.}
\label{delta}
\end{figure}

To determine the mass of the disk when final $\left<e\right>$ is the
same with or without the giant planet, we plotted in the top panel of
Fig. \ref{delta} the values of the final $\left<e\right>$, versus the
total mass in DPs for the cases when the giant is present in the
simulations (blue crosses) and for cases when it is not included (red
crosses). Both red and blue crosses show a progressive increment, as
expected if a population of scatterers is progressively more massive,
i.e., more massive DPs will produce a larger dispersion.  For total DP
masses below 0.096M$_\oplus$, the giant will contribute to maintain a
more ordered disk than if only DPs were present, while above such
limiting mass, the giant substantially contributes to the radial
heating of the disk. This estimate comes from fitting parabolas (shown
as red and blue lines).\\

The bottom panel of Fig. \ref{delta} shows a similar analysis for 
$\left<i\right>$. In this case the regime changes at 0.238M$_\oplus$.\\

\begin{figure}
%\epsscale{.80}
\plotone{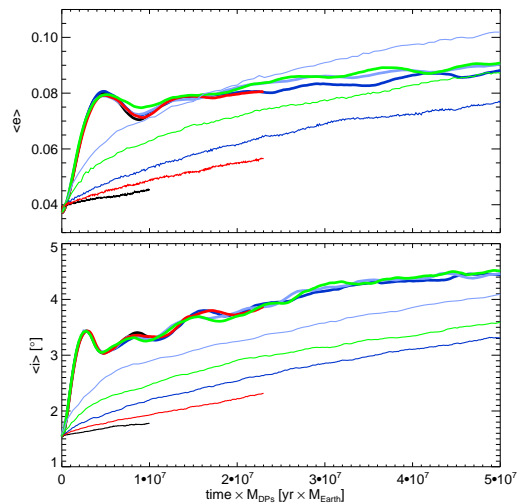}
\caption{Same as in Fig. \ref{massth} but instead of time, we plotted
$\left<e\right>$ and $\left<i\right>$ vs. time multiplied by 
$M_{DPs}$. Thick and thin lines stand for simulations not including, 
and including, the giant planet, respectively.}
\label{eivstm}
\end{figure}

\begin{figure}
%\epsscale{.80}
\plotone{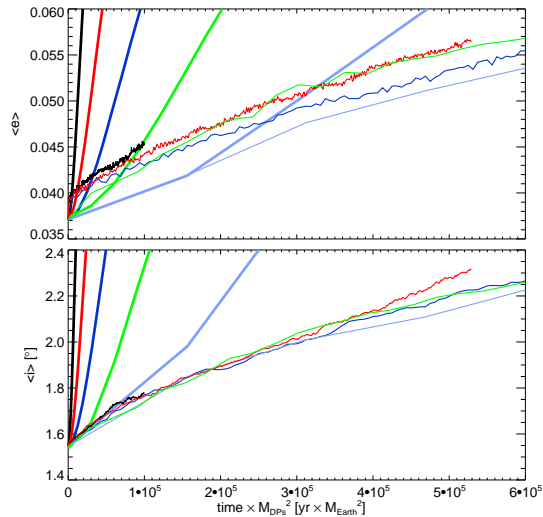}
\caption{Same as in Fig. \ref{massth} but we plotted
$\left<e\right>$ and $\left<i\right>$ vs. time multiplied by 
$M_{DPs}^2$. Again, thick lines are for simulations without the
giant; thin lines for simulations including the giant planet.}
\label{eivstmm}
\end{figure}
 
One would expect the efficiency of the heating to be related to 
the mass of the DPs, and the behavior shown in Fig. \ref{delta}, 
demonstrates that the dependence in mass is different for both 
scenarios (with and without the giant planet). In our simulations 
without the giant planet, we find that the efficiency of the heating 
is proportional to the mass, as expected; in Fig. \ref{eivstm}, we plot 
again the evolution of $\left<e\right>$ and $\left<i\right>$ (for 
the same simulations used in Fig. \ref{massth}), but instead of 
plotting them directly vs. time, we multiply time by the mass of the 
DPs. There we can see that differences are minimal; on the other 
hand, when the giant planet is included, the behavior changes dramatically.
When studying the dependence of the efficiency of the heating with 
the mass, we find that the evolution of the $\left<i\right>$ is 
proportional to the mass of the DPs squared; in Fig. \ref{eivstmm} we 
plot the evolution of $\left<e\right>$ and $\left<i\right>$, in this 
case, vs. time multiplied by the mass of DPs squared, we again see 
how the $\left<i\right>$ curves overlap quite nicely; although the 
overlap of the $\left<e\right>$ is not as good, its behavior is 
still much better represented with the mass squared than with a 
linear dependence.\\

Summarizing, when no giant planet is present, the rate of 
the heating is proportional to the mass of the DPs, however when 
a giant planet is present there is an additional efficiency factor 
and the heating rate is proportional to the mass squared.

\section{Discussion and Conclusions}
\label{conc}

In this work we have explored the evolution of a debris disk that
resembles the cold component of the CKB, but that represents a generic
disk in the sense that it is randomly distributed and formed by small
sized particles. Such disk evolves under the gravitational influence
of 50 DPs which represent the 50 heaviest bodies of three different
mass distributions, with and without the presence of a giant planet,
similar to Neptune.\\
 
Our results have shown that DPs alone induce a progressive stirring in
eccentricities and inclinations of cometary nucleii, proportional to
the mass of the scatterers. Such results may be of importance for very
external debris disks, i.e. disks that lie away from the influence of
any giant planetary body, or for very early protoplanetary disks when
such giant bodies have not yet formed (since it is to be expected that
at least some DP sized objects could form in a short time period,
during the early evolution of protoplanetary systems).\\

On the other hand, when a giant planet is included, we find
that the heating rate of the disk is proportional to the square of the
mass of the DPs. This produces two regimes for the stirring of 
particles in the disk.
In the first one, when the disk mass is below a certain threshold, the
giant planet acts as a stabilizing agent, preventing excessive
dynamical heating of cometary nucleii. Above such threshold, the giant
contributes to the scattering of particles. 
Some consequences can be foreseen over these two
regimes on the stirring and dust formation; the presence of a giant
planet induces stirring for more massive disks and restraining for
light disks (like the Kuiper belt). \\

This tendency would lead us to think that
massive debris disks, with an interior giant planet, would produce 
dust more efficiently in the beginning because of the severe 
increment on the stirring; this is not so straight forward, since 
a giant planet would heat vertically the disk, incrementing rapidly 
its volume, reducing the density, and, with it, reducing the probability 
of collisions among cometary nucleii; whether the dilution of the 
disk or the increment of the stirring will be more important is 
beyond the scope of this paper.
On the other hand, a giant planet
with a lighter disk, could help sustain it to occupy for longer
timescales a given volume, while being stirred by dwarf planets; this
could provide a more sustained mechanism able to produce some dust in
a debris disk for longer timescales. A study on the production of dust
with and without the presence of dwarfs and giant planets is needed
(and is out of the scope of this paper). \\

For the current solar system, the combination between Neptune and the
mass of the Kuiper belt, implies that Neptune acts as a stabilizer
agent for the orbits of the cometary nucleii. Even if a 
larger number of
DPs exist in the Kuiper belt (besides the four currently 
classified as
such by the IAU), it is unlikely that the mass of the Kuiper belt 
would be an
order of magnitude larger than is currently estimated; therefore, it
is hard for them to significantly contribute to the dispersion of the
objects that form the hot classical belt or the scattered
disk. Nonetheless, a detailed study is still necessary to account for
the contribution, in the heating of primordial objects in the cold
classical belt, from known and from yet to be discovered DPs in the
Kuiper belt; this could lead to explain the smooth transition between
the cold and hot classical populations of the Kuiper belt.\\

\acknowledgments

We acknowledge an anonymous referee for a careful and detailed report 
that helped to improve the present work.
We acknowledge 
grant CONACyT Ciencia B\'asica 255167. MAM and BP acknowledge
grant DGAPA-PAPIIT IN114114. AP acknowledges grant
DGAPA-PAPIIT IN109716. We acknowledge the use of the \emph{Atocatl} 
supercomputer at the Instituto de Astronom\'ia
of the Universidad Nacional Aut\'onoma de M\'exico, where most of the 
simulations presented in this work were performed.

\clearpage

%% The following command ends your manuscript. LaTeX will ignore any text


\begin{thebibliography}{}

\bibitem[Aumann et al.(1984)]{1984ApJ...278L..23A} Aumann, H.~H., Beichman, C.~A., Gillett, F.~C., et al.\ 1984, \apjl, 278, L23 
\bibitem[Backman \& Paresce(1993)]{1993prpl.conf.1253B} Backman, D.~E., \& Paresce, F.\ 1993, Protostars and Planets III, 1253 
\bibitem[Bernstein et al.(2004)]{Bernstein04} Bernstein, G.~M., Trilling, D.~E., Allen, R.~L., et al.\ 2004, \aj, 128, 1364 
\bibitem[Chambers(1999)]{Chambers99} Chambers, J.~E.\ 1999, 
\mnras, 304, 793
\bibitem[Chiang et al.(2007)]{2007prpl.conf..895C} Chiang, E., Lithwick, Y., Murray-Clay, R., et al.\ 2007, Protostars and Planets V, 895 
\bibitem[Cotten \& Song(2016)]{2016ApJS..225...15C} Cotten, T.~H., \& Song, I.\ 2016, \apjs, 225, 15 
\bibitem[Dohnanyi(1969)]{Dohnanyi69} Dohnanyi, J.~S.\ 1969, Journal of Geophysical Research, 74, 2531 
\bibitem[Fraser et al.(2008)]{Fraser08a} Fraser, W.~C., Kavelaars, J.~J., Holman, M.~J., et al.\ 2008, \icarus, 195, 827 
\bibitem[Fraser \& Kavelaars(2008)]{Fraser08b} Fraser, W.~C., \& Kavelaars, J.~J.\ 2008, \icarus, 198, 452
\bibitem[Fraser et al.(2014)]{Fraser14} Fraser, W.~C., Brown, M.~E., Morbidelli, A., Parker, A., \& Batygin, K.\ 2014, \apj, 782, 100 
\bibitem[Haisch et al.(2001)]{2001ApJ...553L.153H} Haisch, K.~E., Jr., Lada, E.~A., \& Lada, C.~J.\ 2001, \apjl, 553, L153 
\bibitem[Kenyon \& Bromley(2004)]{Kenyon04} Kenyon, S.~J., \& Bromley, B.~C.\ 2004, \aj, 127, 513 
\bibitem[Kral et al.(2017)]{Kral17} Kral, Q., Clarke, C., \& Wyatt, M.\ 2017, arXiv:1703.08560 
\bibitem[Lee \& Chiang(2016)]{2016ApJ...827..125L} Lee, E.~J., \& Chiang, E.\ 2016, \apj, 827, 125 
\bibitem[Mannings \& Barlow(1998)]{1998ApJ...497..330M} Mannings, V., \& Barlow, M.~J.\ 1998, \apj, 497, 330 
\bibitem[Matthews et al.(2014)]{2014prpl.conf..521M} Matthews, B.~C., Krivov, A.~V., Wyatt, M.~C., Bryden, G., \& Eiroa, C.\ 2014, Protostars and Planets VI, 521 
\bibitem[Morbidelli et al.(2008)]{2008ssbn.book..275M} Morbidelli, A., Levison, H.~F., \& Gomes, R.\ 2008, The Solar System Beyond Neptune, 275 
\bibitem[Moro-Mart{\'{\i}}n et al.(2008)]{Moro08} Moro-Mart{\'{\i}}n, A., Wyatt, M.~C., Malhotra, R., \& Trilling, D.~E.\ 2008, The Solar System Beyond Neptune, 465 
\bibitem[Mouillet et al.(1997)]{1997MNRAS.292..896M} Mouillet, D., Larwood, J.~D., Papaloizou, J.~C.~B., \& Lagrange, A.~M.\ 1997, \mnras, 292, 896 
\bibitem[Mu{\~n}oz-Guti{\'e}rrez et al.(2015)]{Munoz15} Mu{\~n}oz-Guti{\'e}rrez, M.~A., Pichardo, B., Reyes-Ruiz, M., \& Peimbert, A.\ 2015, \apjl, 811, L21 
\bibitem[Mustill \& Wyatt(2009)]{Mustill09} Mustill, A.~J., \& Wyatt, M.~C.\ 2009, \mnras, 399, 1403 
\bibitem[Nesvold \& Kuchner(2015)]{2015ApJ...798...83N} Nesvold, E.~R., \& Kuchner, M.~J.\ 2015, \apj, 798, 83 
\bibitem[Nesvold et al.(2016)]{2016ApJ...826...19N} Nesvold, E.~R., Naoz, S., Vican, L., \& Farr, W.~M.\ 2016, \apj, 826, 19
\bibitem[Nesvold et al.(2017)]{2017arXiv170206578N} Nesvold, E.~R., Naoz, S., \& Fitzgerald, M.\ 2017, arXiv:1702.06578 
\bibitem[Nesvorn{\'y} et al.(2010)]{Nesvorny10} Nesvorn{\'y}, D., Jenniskens, P., Levison, H.~F., et al.\ 2010, \apj, 713, 816 
\bibitem[Nesvorn{\'y}(2015)]{Nesvorny15} Nesvorn{\'y}, D.\ 2015, \aj, 150, 73 
\bibitem[Oudmaijer et al.(1992)]{1992A&AS...96..625O} Oudmaijer, R.~D., van der Veen, W.~E.~C.~J., Waters, L.~B.~F.~M., et al.\ 1992, \aaps, 96, 625 
\bibitem[Smith \& Terrile(1984)]{1984Sci...226.1421S} Smith, B.~A., \& Terrile, R.~J.\ 1984, Science, 226, 1421 
\bibitem[Vitense et al.(2012)]{Vitense12} Vitense, C., Krivov, A.~V., Kobayashi, H., \& L{\"o}hne, T.\ 2012, \aap, 540, A30 
\bibitem[Wyatt(2008)]{2008ARA&A..46..339W} Wyatt, M.~C.\ 2008, \araa, 46, 339 
\bibitem[Wyatt \& Dent(2002)]{2002MNRAS.334..589W} Wyatt, M.~C., \& Dent, W.~R.~F.\ 2002, \mnras, 334, 589 
\bibitem[Wyatt et al.(1999)]{1999ApJ...527..918W} Wyatt, M.~C., Dermott, S.~F., Telesco, C.~M., et al.\ 1999, \apj, 527, 918

\end{thebibliography}
\end{document}